\documentclass[a4paper,11pt]{article}
\pdfoutput=1
\usepackage{jcappub}
\usepackage{epsfig}
\usepackage{amsmath}
\usepackage{amsfonts}
\usepackage{amssymb}
\usepackage{graphicx}
\usepackage{colordvi}

\title{\boldmath Post-Minkowskian Limit and  Gravitational Waves solutions of  Fourth Order Gravity: a complete study}
\vspace{1cm}
\author[a]{A. Stabile,}
\author[b,c,d]{S. Capozziello}

\affiliation[a]{Dipartimento di Ingegneria, Universit\`{a} del Sannio, Corso Garibaldi, 107 - 82100 Benevento, Italy}
\affiliation[b]{Dipartimento di Fisica, Universit\`{a} di Napoli "Federico II", Complesso Universitario di Monte Sant'Angelo, Edificio G, Via Cinthia, I-80126, Napoli, Italy}
\affiliation[c]{INFN Sezione di Napoli, Complesso Universitario  di Monte Sant'Angelo, Edificio G, Via Cinthia, I-80126, Napoli, Italy}
\affiliation[d]{Gran Sasso Science Institute (INFN), Viale F. Crispi, 7, I-67100. L'Aquila, Italy.}

\emailAdd{arturo.stabile@gmail.com}
\emailAdd{capozziello@na.infn.it}

\abstract {The post-Minkowskian limit and gravitational wave solutions for general  fourth-order gravity  theories are discussed. Specifically, we consider a Lagrangian with a generic function of curvature invariants  $f(R, R_{\alpha\beta}R^{\alpha\beta}, R_{\alpha\beta\gamma\delta}R^{\alpha\beta\gamma\delta})$. It is well known that when dealing with General Relativity such an approach provides massless spin-two waves as propagating degree of freedom of the gravitational field while this theory implies other additional propagating modes in the gravity spectra. We show that, in general,   fourth order gravity,  besides the standard massless graviton is characterized by two further massive modes with a finite-distance interaction. We find out the most general gravitational wave solutions in terms of Green functions in  vacuum and in presence of  matter sources. If an electromagnetic source is chosen, only the modes induced by $R_{\alpha\beta}R^{\alpha\beta}$ are present, otherwise, for any $f(R)$ gravity model, we have the complete analogy with tensor modes of  General Relativity. Polarizations and helicity states are classified  in  the hypothesis of plane wave.}

\begin{document}
\maketitle
\flushbottom

\section{Introduction}

Identifying the correct theory of gravity is a crucial issue of modern physics due to the fact that General Relativity, in its standard formulation, presents shortcomings at ultraviolet and infrared limits. For the first issue, we need a theory that  should deal with gravity under the same standard of the other fundamental  interactions (Quantum Gravity) \cite{quantum}. In the other case, modifications of gravity are required to deal with the vast  phenomenology coming from astrophysics and cosmology, generally addressed as dark matter and dark energy issues \cite{dark1,dark2}. This matter is rather controversial due to the fact that the ambiguity comes out from the fact that phenomenology could be explained successfully both considering new material ingredients (dark matter particles  addressing the problem of structure formation and  light scalar fields giving rise to the acceleration of the Hubble fluid) or modifying gravity that, at scales larger than Solar System, could behave in different way with respect to the weak field limit related to General Relativity \cite{dark3,dark4,dark5, schmidt,schmidt2}. Furthermore, there are very few investigations and experimental constraints  probing  the  gravitational field in  very strong regimes. Several times, extrapolations of General Relativity are simply assumed without considering corrections and alternatives that could strongly affect theoretical and experimental results.

With this situation in mind, it is urgent to find out some {\it experimentum crucis} or some test bed capable of discriminating among concurring gravitational theories \cite{new_janis,stabile_scelza,stabile_brothers,PRD_sake}, that, in any case, should reproduce the well-founded theoretical and experimental results of General Relativity. At astrophysical level, discriminations could come from anomalous stellar systems whose structures and parameters do not find room in the constraints and limits imposed by General Relativity. For example, extremely massive neutron stars, magnetars or compact objects like quark stars could be independent {\it signatures} for modified theories of gravity considered as extensions of General Relativity in the strong field regime \cite{neutron1,neutron2}.

Besides, discrimination could happen in the realm of gravitational wave physics. This sector of physics, practically unexplored from the point of view of modified gravity,  deserves a lot of attention since the large part of efforts has been devoted to the study of gravitational radiation in the realm of General Relativity discarding the fact that modified gravity presents  a huge amount of new phenomenology and features. For example, only  General Relativity  strictly forecasts massless gravitons with two polarizations. In general, modified gravity and, in particular Extended Gravity, allows also massive and ghost modes and then further polarizations \cite{greci1,greci2}. Specifically, several  authors refer this issue to the Fierz-Pauli linearized  analysis of massive gravity that leads to the so called van Dam-Veltman-Zakharov (vDVZ) discontinuity. In such a case, we are in presence of the Boulware-Deser ghosts. Such  anomalies can be cured through the Vainshtein mechanism. A detailed review of these arguments can be found in \cite{derham}.

However, the possibility of these massive modes are  studied from a theoretical point of view but practically ignored from the experimental point of view due to the enormous difficulties related to the detection of gravitational waves. However, the forthcoming experimental facilities like VIRGO (after the Virgo supercluster of galaxies)\cite{virgo}, LIGO (Laser Interferometer Gravitational-Wave Observatory \cite{ligo}) collaboration, LISA (aser Interferometer Space Antenna \cite{lisa}) etc. could be suitable, in principle,  for detecting these further modes. 

In this paper, we propose a systematic study of gravitational wave solutions in theories where generic  functions of curvature invariants are considered generalizing the first partial outcome in the only $f(R)$ framework \cite{minko}. These are a straightforward generalization of $f(R)$ gravity where the degrees of freedom, related to the curvature invariants, is considered. However, we have to say that we are not considering  $\Box R$ and similar terms where derivative of curvature invariants appear.  We are also not considering the parity-odd Chern-Simons invariant \cite{chern} that enters at the same order in curvatures and derivatives. Here, we are taking into account only fourth order theories of gravity where derivatives of metric tensor $g_{\mu\nu}$ appear up to the fourth ones. It is interesting to see that relaxing the hypothesis that gravitational interaction is  derived only from the Hilbert-Einstein action, linear in the Ricci curvature scalar $R$, further gravitational  modes, polarizations and helicity states come out. This new features are directly derived from the post-Minkowskian limit of the theory and points out a new rich phenomenology that deserves investigation in view of possible future detection of gravitational waves.
The classification of these  modes, implying new massive  and polarization states, is the same of Wigner's little group $E(2)$ \cite{wigner1,wigner2} earlier explored in literature by Eardley et al. \cite{eardley}. Besides, $f(R)$ theories of gravity have been very well explored in this sense. For example, it is well known that they are equivalent to scalar-tensor theories of gravity following the Bergmann-Wagoner formulation  \cite{bergmann,wagoner} which generalizes the Brans-Dicke theory.  Furthermore, kinematics of gravitational waves in $f(R)$ gravity have been extensively studied by Berry and Gair which confronted gravitational radiation with Solar System tests \cite{gair}. It is important to stress that being $f(R)$ gravity equivalent to a particular class of scalar-tensor gravity theories, the gravitational wave content has  already been characterized in both representations \cite{lang}. Also a comparison between the two frameworks (the Einstein and Jordan frame) has been analyzed in the weak field limit \cite{conf_trans,ST_fR}. Finally, gravitational wave kinematics and dynamics in strong field regime and around the Minkowski background have been investigated in detail \cite{stein}.

This paper is  organized as follows. In Section \ref{FOG} we report briefly the field equations of  fourth order gravity. In  Section \ref{minkow_sec}, we discuss the post-Minkowskian limit and the linearized field equations while, in Section \ref{PM_solutions}, the gravitational wave solutions are reported.  Section \ref{polar} is devoted to the  discussion of  all  possible   polarization and helicity states of the wave solutions. Conclusions are reported in Section \ref{conclusions}.

\section{The field equations of Fourth Order Gravity}\label{FOG}

The most general class  of gravitational  theories involving curvature invariants in four dimensions is given by the action

\begin{eqnarray}\label{FOGaction}\mathcal{A}=\int d^{4}x\sqrt{-g}\biggl[f(X,Y,Z)+\mathcal{X}\mathcal{L}_m\biggr]
\end{eqnarray}
where $f$ is an unspecified function of curvature invariants $X\,=\,R$,
$Y\,=\,R_{\alpha\beta}R^{\alpha\beta}$, and $Z\,=\,\,R_{\alpha\beta\gamma\delta}R^{\alpha\beta\gamma\delta}$. The term $\mathcal{L}_m$ is the minimally coupled
ordinary matter contribution. In the metric approach, the field
equations are obtained by varying (\ref{FOGaction}) with respect
to $g_{\mu\nu}$. We get the fourth-order differential  equations

\begin{eqnarray}\label{fieldequationFOG}
H_{\mu\nu}\,=&&f_XR_{\mu\nu}-\frac{f}{2}g_{\mu\nu}-f_{X;\mu\nu}+g_{\mu\nu}\Box
f_X+2f_Y{R_\mu}^\alpha
R_{\alpha\nu}-2[f_Y{R^\alpha}_{(\mu}]_{;\nu)\alpha}+\Box[f_YR_{\mu\nu}]\nonumber\\\\&&+[f_YR_{\alpha\beta}]^{;\alpha\beta}g_{\mu\nu}+2f_ZR_{\mu\alpha
\beta\gamma}{R_{\nu}}^{\alpha\beta\gamma}
-4[f_Z{{R_\mu}^{\alpha\beta}}_\nu]_{;\alpha\beta}\,=\,
\mathcal{X}\,T_{\mu\nu}\nonumber
\end{eqnarray}
where
$T_{\mu\nu}\,=\,-\frac{1}{\sqrt{-g}}\frac{\delta(\sqrt{-g}\mathcal{L}_m)}{\delta
g^{\mu\nu}}$ is the the energy-momentum tensor of matter,
$f_X\,=\,\frac{\partial f}{\partial X}$, $f_Y\,=\,\frac{\partial f}{\partial Y}$,
$f_Z\,=\,\frac{\partial f}{\partial Z}$, $\Box={{}_{;\sigma}}^{;\sigma}$, and
$\mathcal{X}\,=\,8\pi G$\footnote{Here we use the convention
$c\,=\,1$.}. The conventions for Ricci's tensor is
$R_{\mu\nu}={R^\sigma}_{\mu\sigma\nu}$ and  for the Riemann
tensor is
${R^\alpha}_{\beta\mu\nu}=\Gamma^\alpha_{\beta\nu,\mu}+...$. The
affinities are the usual Christoffel's symbols of the metric:
$\Gamma^\mu_{\alpha\beta}=\frac{1}{2}g^{\mu\sigma}(g_{\alpha\sigma,\beta}+g_{\beta\sigma,\alpha}-g_{\alpha\beta,\sigma})$. The adopted signature is $(+---)$ (we follow the conventions by Landau and Lifshitz \cite{landau}). The trace of field Eqs.
(\ref{fieldequationFOG}) is the following

\begin{eqnarray}\label{tracefieldequationFOG}
H\,=\,f_XX+2f_YY+2f_ZZ-2f+\Box[3
f_X+f_YX]+2[(f_Y+2f_Z)R^{\alpha\beta}]_{;\alpha\beta}\,=\,\mathcal{X}\,T
\end{eqnarray}
where $T\,=\,T^{\sigma}_{\,\,\,\,\,\sigma}$ is the trace of
energy-momentum tensor and $H\,=\,H^{\sigma}_{\,\,\,\,\,\sigma}$.

Some authors considered a linear Lagrangian containing not only $X$, $Y$ and $Z$ but also the first power of curvature invariants $\Box R$ and ${R^{\alpha\beta}}_{;\alpha\beta}$. Such a  choice is justified because all curvature invariants have the same dimension ($L^{-2}$) \cite{santos}. Furthermore, this dependence on the  two last invariants is only formal, since from the contracted Bianchi identity ($2{R^{\alpha\beta}}_{;\alpha\beta}-\Box R\,=\,0$) we have only one independent invariant. In any linear theory of gravity (the function $f$ is linear) the terms $\Box R$ and ${R^{\alpha\beta}}_{;\alpha\beta}$ give us no contribution to the field equations, because they are four-divergences. However if we consider a function of $\Box R$ or ${R^{\alpha\beta}}_{;\alpha\beta}$ by varying the action,  we still have  four-divergences but we would have the contributions of sixth order differential terms. As said in the Introduction, in  this paper,  we consider only  fourth order differential field equations. This means that Action (\ref{FOGaction}) is a fourth-order theory and no higher-order terms in derivatives are taken into account. However, as we will see below, one needs only two of the three curvature invariants, due to the Gauss-Bonnet topological invariant which fixes a constraint among the curvature terms.

\section{The Post Minkowskian Limit}\label{minkow_sec}

Any theory of gravity has to be discussed in the 
\emph{weak field limit approximation}. This \emph{prescription} is
needed to test if the given theory is consistent with the
well-established Newtonian theory and with the Special Relativity
as soon as the the gravitational field is weak or is almost null.
Both requirements are fulfilled by General Relativity  and then they can be
considered two possible paradigms to confront a given theory, at
least in the weak field limit, with the  General Relativity  itself. The Newtonian limit of $f(R)$-gravity and
$f(R,\,R_{\alpha\beta}R^{\alpha\beta},\,R_{\alpha\beta\gamma\delta}R^{\alpha\beta\gamma\delta})$-gravity can be  investigated always remaining in the Jordan
frame \cite{noi-newt,noi-newt_2,dirk,mio_1,mio_2,FOG_ST} while a preliminary study of the post-Minkoskian limit for  the  $f(R)$ gravity is  provided in Ref. \cite{minko_FOG}. Here we want to derive the
post-Minkowskian limit of $f(R,\,R_{\alpha\beta}R^{\alpha\beta},\,R_{\alpha\beta\gamma\delta}R^{\alpha\beta\gamma\delta})$ gravity, that is for generic  fourth-order theory of gravity, with the  aim to investigate the gravitational radiation.

The post-Minkowskian limit of any theory of  gravity arises when
the regime of small field is considered without any prescription
on the propagation of the field. This case has to be clearly
distinguished with respect to the Newtonian limit which,
differently, requires both the small velocity and the weak field
approximations. Often, in literature, such a distinction is not
clearly remarked and several cases of pathological analysis can be
accounted. The post-Minkowskian limit of General Relativity gives rise to massless
gravitational waves and reproduces  the Special Relativity. An analogous study can be pursued
considering, instead of the Hilbert-Einstein Lagrangian, linear in
the Ricci scalar $R$, a most general function $f$ of curvature
invariants.

However, working with post-Minkowskian limit, one has to be extremely careful with the issue  whether or not there are  ghost modes that could render the theory meaningless. A standard way to deal with ghost modes is to consider any alternative  theory as an effective field theory  close to General Relativity, since  it  is well known that  General Relativity   has  no ghosts. In fact, effective field theory has a restricted regime of validity, and the ghost modes only appear once one leaves such a regime. See, for examples   \cite{dyda, ricciardi} where such a problem is discussed for the Chern-Simons and $f(R)$ theories respectively. Essentially, the approach consists in  decoupling the modes in the weak limit. In \cite{stein},  it was shown that in the decoupling limit, and approaching the asymptotically flat region far from any matter sources, a very generic class of theories with higher curvature invariants leads  exactly to the  same gravitational wave modes  as in General Relativity (see for example \cite{bamba} for the case of $f(T)$ teleparallel gravity).

Another  question is related  the Cauchy problem of the theory, and in particular with its well-formulation and  well-posed initial value formulation. For $f(R)$ gravity, this problem has been highly debated and discussed \cite{faraoni,vignolo1,vignolo2,r2}. In any case, considering theories  with higher derivatives, the Cauchy problem has to be treated with care   because the related Hamiltonians could   suffer, in general,  with the Ostrogradski instability (see \cite{querella} for a detailed study). In particular,  the existence of a well-posed initial value problem for higher-order theories of gravity  involves the  introduction  of auxiliary degrees of freedom to reduce the derivative order. In some sense, the mechanism is the same acting in conformal transformations where the further degrees of freedom in the Jordan frame are disentangled in scalar fields minimally coupled to gravity in the Einstein frame \cite{greci2}.  This fact allows to evaluate how many propagating dynamical degrees of freedom are present in the theory. Alternatively, treating the theory as an effective field theory  can contribute to  address the question of the existence of a well-posed initial value formulation \cite{delsate}.

In order to perform the post-Minkowskian limit of field
equations, one has to perturb  Eqs. (\ref{fieldequationFOG})  on the
Minkowski background $\eta_{\mu\nu}$. In such a case,  we obtain

\begin{equation}\label{PM_me}
g_{\mu\nu}\,=\,\eta_{\mu\nu}+h_{\mu\nu}
\end{equation}
with $h_{\mu\nu}$ small ($\mathcal{O}({h_{\mu\nu}}^2)\ll 1$). Then the curvature invariants $X$, $Y$,
$Z$ become

\begin{eqnarray}
\left\{\begin{array}{ll}
X\,\sim\,X^{(1)}+\mathcal{O}({h_{\mu\nu}}^2)\\\\
Y\,\sim\,Y^{(2)}+\mathcal{O}({h_{\mu\nu}}^3)\\\\
Z\,\sim\,Z^{(2)}+\mathcal{O}({h_{\mu\nu}}^3)
\end{array}\right.
\end{eqnarray}
and the function $f$ can be developed as

\begin{eqnarray}
f(X,Y,Z)\,\sim\,f(0)+f_X(0)X^{(1)}+\frac{1}{2}f_{XX}(0){X^{(1)}}^2+f_Y(0)Y^{(2)}+f_Z(0)Z^{(2)}+\mathcal{O}({h_{\mu\nu}}^3)
\end{eqnarray}
Analogous relations for partial derivatives of $f$ are
obtained. From lowest order of
field Eqs. (\ref{fieldequationFOG}) and
(\ref{tracefieldequationFOG}) we have the condition $f(0)\,=\,0$,
while at $\mathcal{O}(1)$ - order,  we have\footnote{We are using the properties:
$2{R_{\alpha\beta}}^{;\alpha\beta}-\Box R\,=\,0$ and
${{R_\mu}^{\alpha\beta}}_{\nu;\alpha\beta}\,=\,{{R_\mu}^\alpha}_{;\nu\alpha}-\Box
R_{\mu\nu}$.}

\begin{eqnarray}\label{PMfieldequationFOG}
\begin{array}{ll}
f_X(0)R^{(1)}_{\mu\nu}+[f_Y(0)+4f_Z(0)]\Box_\eta
R^{(1)}_{\mu\nu}
-\frac{f_X(0)}{2}X^{(1)}\eta_{\mu\nu}+[f_{XX}(0)+\frac{f_Y(0)}{2}]\eta_{\mu\nu}\Box_\eta
X^{(1)}\\\\\,\,\,\,\,\,\,\,\,\,\,\,\,\,\,\,\,\,\,\,\,\,\,\,\,\,-f_{XX}(0){X^{(1)}}_{,\mu\nu}-[f_Y(0)+4f_Z(0)]
R^{(1)\alpha}_{\,\,\,\,\,\,\,\,\,\,\,\mu,\nu\alpha}-f_Y(0)
R^{(1)\alpha}_{\,\,\,\,\,\,\,\,\,\,\,\nu,\mu\alpha}\,=\,\mathcal{X}\,T^{(0)}_{\mu\nu}\\\\
-f_X(0)X^{(1)}+[3f_{XX}(0)+2f_Y(0)+2f_Z(0)]\Box_\eta
X^{(1)}\,=\,\mathcal{X}\,T^{(0)}
\end{array}
\end{eqnarray}
where $\Box_\eta$ is the d'Alembert operator in the flat space.
$T_{\mu\nu}$ is fixed at zero-order in (\ref{PMfieldequationFOG})
since, in this perturbation scheme, the first order on Minkowski
space has to be connected with the zero order, with respect to $h_{\mu\nu}$, of the standard
matter energy momentum tensor.
 This means that  $T_{\mu\nu}$ is indipendent of $h_{\mu\nu}$, and satisfies the standard conservation conditions $T^{\mu\nu}_{\,\,\,\,\,\,\,,\mu}\,=\,0$. By introducing the quantities

\begin{eqnarray}\label{mass_definition}
\begin{array}{ll}
{m_1}^2\,\doteq\,-\frac{f_X(0)}{3f_{XX}(0)+2f_Y(0)+2f_Z(0)}\\\\
{m_2}^2\,\doteq\,\frac{f_X(0)}{f_Y(0)+4f_Z(0)}
\end{array}
\end{eqnarray}
we get two differential equations for curvature invariant
$X^{(1)}$ and Ricci tensor $R^{(1)}_{\mu\nu}$\footnote{We set $f_{X}\,=\,1$ \emph{i.e.} $G\,\rightarrow\,f_{X}(0)\,G$.}

\begin{eqnarray}\label{PMfieldequationFOG2}
\begin{array}{ll}
(\Box_\eta+{m_2}^2)R^{(1)}_{\mu\nu}-\biggl[\frac{{m_1}^2-{m_2}^2}{3{m_1}^2}\,
\partial^2_{\mu\nu}+\eta_{\mu\nu}\biggl(\frac{{m_2}^2}{2}+\frac{{m_1}^2+2{m_2}^2}{6{m_1}^2}\Box_\eta\biggr)\biggr]
X^{(1)}\,=\,{m_2}^2\,\mathcal{X}\,T^{(0)}_{\mu\nu}\\\\
(\Box_\eta+{m_1}^2)X^{(1)}\,=\,-{m_1}^2\,\mathcal{X}\,T^{(0)}
\end{array}
\end{eqnarray}
We note that in the case of $f(X)$-theory we obtain only a
massive mode (with mass $m_1$) of Ricci scalar ($X$). In fact if $f_Y\,=\,f_Z\,=\,0$ from the mass definition (\ref{mass_definition}) ${m_1}^2\,\rightarrow\,(-3f_{XX}(0))^{-1}$ and ${m_2}^2\,\rightarrow\,\infty$ we recover the equations of $f(X)$-gravity \cite{minko_FOG}

\begin{eqnarray}\label{PMfieldequationFR}
\begin{array}{ll}
R^{(1)}_{\mu\nu}-\frac{X^{(1)}}{2}\eta_{\mu\nu}+\frac{\partial^2_{\mu\nu}-\eta_{\mu\nu}\Box_\eta}{3{m_1}^2}\,
X^{(1)}\,=\,\mathcal{X}\,T^{(0)}_{\mu\nu}\\\\
(\Box_\eta+{m_1}^2)X^{(1)}\,=\,-\,{m_1}^2\mathcal{X}\,T^{(0)}
\end{array}
\end{eqnarray}
while $f(X,Y,Z)$-theory we have an additional massive
propagation (with mass $m_2$) of Ricci tensor. Finally in the case of $f\,\rightarrow\,X$ also ${m_1}^2\,\rightarrow\,\infty$ and we recover the General Relativity.

A first consideration regarding  the masses (\ref{mass_definition}) induced by $f(X,Y,Z)$-gravity is necessary at this point. The second mass $m_2$ is originated by the presence, in the
Lagrangian, of Ricci and Riemann tensor square, but also a theory
containing only Ricci tensor square gives rise to  the same outcome.
Obviously the same is valid also with the Riemann tensor square
alone. Then  such a modification of theory enables a massive
propagation of Ricci Tensor and, as it is well known in the
literature, a substitution of the Ricci scalar with any function of the 
Ricci scalar enables a massive propagation of Ricci scalar. We
can conclude that a Lagrangian containing any
function of only Ricci scalar and Ricci tensor square is not
restrictive. This result is coming from  the Gauss - Bonnet topological 
invariant $G_{GB}$ defined as $G_{GB}\,=\,X^2-4Y+Z$
 \cite{dewitt_book}. In fact by applying the variation with respect to the metric tensor $g_{\mu\nu}$ in 4 dimensions to the  quantity  $\int d^4x\sqrt{-g}\,G_{GB}=0$, we have

\begin{eqnarray}\label{fieldequationGB_0}
\delta\,\int d^4x\sqrt{-g}\,G_{GB}\,=\,\int d^4x\sqrt{-g}\,(H^{X^2}_{\mu\nu}-4H^{Y}_{\mu\nu}+H^{Z}_{\mu\nu})\delta g^{\mu\nu}\,=\,\int d^4x\sqrt{-g}\,H^{GB}_{\mu\nu}\,\delta g^{\mu\nu}
\end{eqnarray}
where $H^{X^2}_{\mu\nu}\,=\,\frac{1}{\sqrt{-g}}\frac{\delta(\sqrt{-g}\,X^2)}{\delta g^{\mu\nu}}$, $H^Y_{\mu\nu}\,=\,\frac{1}{\sqrt{-g}}\frac{\delta(\sqrt{-g}\,Y)}{\delta g^{\mu\nu}}$ and $H^Z_{\mu\nu}\,=\,\frac{1}{\sqrt{-g}}\frac{\delta(\sqrt{-g}\,Z)}{\delta g^{\mu\nu}}$. In four dimensions we have that $H^{GB}_{\mu\nu}$ is identically zero. For this reason the variation of the Gauss-Bonnet invariant generates a dimension-dependent identity, that is

\begin{eqnarray}\label{fieldequationGB}
H^{X^2}_{\mu\nu}-4H^{Y}_{\mu\nu}+H^{Z}_{\mu\nu}\,=\,0
\end{eqnarray}
By substituting  condition (\ref{fieldequationGB})  in Eqs. (\ref{fieldequationFOG}), at  post-Minkowskian level,     we find the same Eqs. (\ref{PMfieldequationFOG2}) with a redefinition of  masses (\ref{mass_definition}). In the weak field limit approximation,  we can consider as Lagrangian in the action (\ref{FOGaction}),  the quantity \cite{mio_2}

\begin{eqnarray}
f(X,Y,Z)\,=\,a\,X+b\,X^2+c\,Y
\end{eqnarray}
Then the masses (\ref{mass_definition}) become

\begin{eqnarray}\label{mass_definition_NL}
\begin{array}{ll}
{m_1}^2\,=\,-\frac{a}{2(3b+c)}
\\\\
{m_2}^2\,=\,\frac{a}{c}
\end{array}
\end{eqnarray}
These  masses have real values  if  the conditions $a\,>\,0$, $b\,<\,0$ and $0\,<\,c\,<\,-3b$ hold. This fact is extremely important in order to establishing a no-ghost constraint on such a theory. For gravity theories  explicitly containing the Gauss-Bonnet term  in the post-Newtonian limit see \cite{felix}.

\section{Gravitational wave solutions}\label{PM_solutions}

Once developed the post-Minkowskian limit, one can search for gravitational wave solutions. The general solution of field equations (\ref{PMfieldequationFOG2}) is given by 

\begin{eqnarray}
h_{\mu\nu}\,=\,\mathfrak{h}_{\mu\nu}+\mathsf{h}_{\mu\nu}
\end{eqnarray}
where $\mathfrak{h}_{\mu\nu}$ is the (homogeneous) solution in the vacuum and $\mathsf{h}_{\mu\nu}$ is the (particular) one in the matter. First we try to find the solution $\mathsf{h}_{\mu\nu}$. From the second line of (\ref{PMfieldequationFOG2}), by introducing the Green function $\mathcal{G}_{KG,i}(x,x')$ of Klein-Gordon field defined as follows

\begin{eqnarray}
(\Box_\eta+{m_i}^2)\,\mathcal{G}_{KG,i}(x,x')\,=\,\delta^{(4)}(x-x'),\,\,\,\,\,\,\text{with}\,\,i\,=\,1,2
\end{eqnarray}
where $\delta^{(4)}(x-x')$ is the Dirac delta function in four dimensions, we find

\begin{eqnarray}\label{solutionRicciscalar}
X^{(1)}\,=\,-{m_1}^2\mathcal{X}\int d^4x'\,\mathcal{G}_{KG,1}(x,x')\,T^{(0)}(x')
\end{eqnarray}
where $x\,=\,x^\mu\,=\,(t,\,\mathbf{x})\,=\,(t,\,x^1,\,x^2,\,x^3)$. The first line of (\ref{PMfieldequationFOG2}) can be recast as follows

\begin{eqnarray}\label{PMfieldequationFOG3}
(\Box_\eta+{m_2}^2)R^{(1)}_{\mu\nu}\,=&&\mathcal{X}\biggl[{m_2}^2\,T^{(0)}
_{\mu\nu}-\frac{{m_1}^2+2{m_2}^2}{6}\eta_{\mu\nu}T^{(0)}\biggl]
\nonumber\\\\&&-\frac{({m_1}^2-{m_2}^2)
\mathcal{X}}
{3}\,\int d^4x'\biggl[
\partial^2_{\mu\nu}-\frac{{m_1}^2}{2}\eta_{\mu\nu}\bigg]\,\mathcal{G}_{KG,1}(x,x')\,T^{(0)}(x')\nonumber
\end{eqnarray}
so the solution for the Ricci tensor is obtained

\begin{eqnarray}\label{PMfieldequationFOG4}
R^{(1)}_{\mu\nu}\,=&&\mathcal{X}\int d^4x'\,\mathcal{G}_{KG,2}(x,x')\biggl[{m_2}^2\,T^{(0)}
_{\mu\nu}(x')-\frac{{m_1}^2+2{m_2}^2}{6}\eta_{\mu\nu}T^{(0)}(x')\biggl]\nonumber\\\\&&\nonumber
-\frac{({m_1}^2-{m_2}^2)\mathcal{X}}
{3}\,\int d^4x'\,d^4x''\,\mathcal{G}_{KG,2}(x,x')\biggl[
\partial^2_{\mu'\nu'}-\frac{{m_1}^2}{2}\eta_{\mu\nu}\bigg]\,\mathcal{G}_{KG,1}(x',x'')\,T^{(0)}(x'')
\end{eqnarray}
The Ricci tensor, in terms of the metric (\ref{PM_me}), is given by

\begin{equation}\label{Riccitensor}
R^{(1)}_{\mu\nu}\,=\,h^\sigma_{\,\,\,(\mu,\nu)\sigma}-\frac{1}{2}\Box_\eta\,
h_{\mu\nu}-\frac{1}{2}h_{,\mu\nu}
\end{equation}
where $h\,=\,{h^\sigma}_{\,\sigma}$. Since we can use the harmonic gauge condition $g^{\rho\sigma}\Gamma^\alpha_{\,\,\,\rho\sigma}\,=\,0$, we set $h_{\mu\sigma}^{\,\,\,\,\,\,\,,\sigma}-1/2\,h_{\,,\mu}\,=\,0$, then the Ricci tensor becomes $R^{(1)}_{\mu\nu}\,=\,-\frac{1}{2}\Box_\eta\,
h_{\mu\nu}$. The solution of Eq.(\ref{PMfieldequationFOG4}) is

\begin{eqnarray}\label{PMfieldequationsolution}
\mathsf{h}_{\mu\nu}\,=&&\frac{2(
{m_1}^2-{m_2}^2)\mathcal{X}}
{3}\,\int d^4x'\,d^4x''\,d^4x'''\,\mathcal{G}_{GR}(x,x')\,\mathcal{G}_{KG,2}(x',x'')
\nonumber\\\nonumber\\&&
\qquad\qquad\times\biggl[\partial^2_{\mu''\nu''}-\frac{{m_1}^2}{2}\eta_{\mu\nu}\bigg]\,\mathcal{G}_{KG,1}(x'',x''')\,T^{(0)}(x''')
\\\nonumber\\&&-2\mathcal{X}\int d^4x'\,d^4x''\,\mathcal{G}_{GR}(x,x')\,\mathcal{G}_{KG,2}(x',x'')\biggl[{m_2}^2\,
T^{(0)}_{\mu\nu}(x'')-\frac{{m_1}^2+2{m_2}^2}{6}\eta_{\mu\nu}T^{(0)}(x'')\biggl]\nonumber
\end{eqnarray}
where $\mathcal{G}_{GR}(x,x')$ is the Green function defined as

\begin{eqnarray}
\Box_\eta\,\mathcal{G}_{GR}(x,x')\,=\,\delta^{(4)}(x-x')
\end{eqnarray}
Considering the expressions of the Green functions $\mathcal{G}_{GR}(x,x')$ and $\mathcal{G}_{KG,1,2}(x,x')$ in terms of plane waves it is possible to rewrite the solution (\ref{PMfieldequationsolution}) as follows

\begin{eqnarray}\label{PMfieldequationsolution3}
\begin{array}{ll}
\mathsf{h}_{\mu\nu}(x)\,=\,\mathcal{X}\,\int d^4x'\biggl[\mathcal{Z}_2(x,x')\,
T^{(0)}_{\mu\nu}(x')+\biggl(\mathcal{Z}(x,x')\,\eta_{\mu\nu}+\mathcal{Z}_{\mu\nu}(x,x')\biggr)\,T^{(0)}(x')\biggr]\\\\
\mathsf{h}(x)\,=\,\mathcal{X}\,\int d^4x'\,\mathcal{Z}_1(x,x')\,T^{(0)}(x')
\end{array}
\end{eqnarray}
where

\begin{eqnarray}\label{greenfunctions_2}
\begin{array}{ll}
\mathcal{Z}_s(x,x')\,=\,2{m_s}^2\,(-1)^{1+s}\,\int\frac{d^4k}{(2\pi)^4}\frac{e^{\mathfrak{j}\,k(x-x')}}{k^2(k^2-{m_s}^2)}\\\\
\mathcal{Z}(x,x')\,=\,\int \frac{d^4k}{(2\pi)^4}\frac{({m_1}^2+2{m_2}^2)k^2-3{m_1}^2{m_2}^2}{3\,k^2(k^2-{m_1}^2)(k^2-{m_2}^2)}\,e^{\mathfrak{j}\,k(x-x')}\\\\
\mathcal{Z}_{\mu\nu}(x,x')\,=\,\int \frac{d^4k}{(2\pi)^4}\frac{2({m_1}^2-{m_2}^2)k_\mu k_\nu}{3\,k^2(k^2-{m_1}^2)(k^2-{m_2}^2)}\,e^{\mathfrak{j}\,k(x-x')}
\end{array}
\end{eqnarray}
$k\,=\,k^\mu\,=\,(\omega,\,\mathbf{k})\,=\,(\omega,\,k^1,\,k^2,\,k^3)$, $k\,x\,=\,k_\sigma x^\sigma\,=\,\omega\,t\,-\,\mathbf{k}\cdot\mathbf{x}$, $k^2\,=\,k_\sigma k^\sigma\,=\,\omega^2-|\mathbf{k}|^2$ and $s\,=\,1,\,2$. We note that in the case $f\,\rightarrow\,X$ (from the mass definitions (\ref{mass_definition}) we have $m_1,\,m_2\,\rightarrow\,\infty$) we find $\mathcal{Z}_s(k)\,\rightarrow\,2(-1)^sk^{-2}$, $\mathcal{Z}(k)\,\rightarrow\,-k^{-2}$, $\mathcal{Z}_{\mu\nu}(k)\,\rightarrow\,0$ and the solutions (\ref{PMfieldequationsolution3}) become those of General Relativity: 
\begin{equation}
\mathsf{h}_{\mu\nu}(x)\,=\,-2\mathcal{X}\,\int d^4x'\mathcal{G}_{GR}(x,x')\,S^{(0)}_{\mu\nu}(x')\,,
\end{equation} 
where $S^{(0)}_{\mu\nu}\,=\,T^{(0)}_{\mu\nu}-\eta_{\mu\nu}\,T^{(0)}/2$.

Finally we can recast the propagators (\ref{greenfunctions_2}) in terms of the Green functions $\mathcal{G}_{GR}(x,x')$ and $\mathcal{G}_{KG,s}(x,x')$ and we can see immediately the propagation of the massless interaction and of two massive interactions. We get

\begin{eqnarray}\label{greenfunctions_3}
\begin{array}{ll}
\mathcal{Z}_s(x,x')\,=\,2(-1)^{1+s}\,\biggl[\mathcal{G}_{GR}(x,x')-\mathcal{G}_{KG,s}(x,x')\biggr]\\\\
\mathcal{Z}(x,x')\,=\,\mathcal{G}_{GR}(x,x')-\frac{1}{3}\,\mathcal{G}_{KG,1}(x,x')-\frac{2}{3}\mathcal{G}_{KG,2}(x,x')\\\\
\mathcal{Z}_{\mu\nu}(x,x')\,=\,\frac{2}{3}\partial^2_{\mu\nu}\biggl[\frac{{m_1}^2-{m_2}^2}{{m_1}^2{m_2}^2}\,\mathcal{G}_{GR}(x,x')+\frac{\mathcal{G}_{KG,1}(x,x')}{{m_1}^2}-\frac{\mathcal{G}_{KG,2}(x,x')}{{m_2}^2}\biggr]
\end{array}
\end{eqnarray}
It is important to stress that  the limit $m\rightarrow 0$ could become  problematic for the above prapagators since the theory appears strongly coupled in this limit.  However the source term can be assumed decoupled as soon as $m\rightarrow 0$. In such a case the problem is solved. See \cite{derham} for a discussion.

A very interesting feature is achieved when we consider a traceless source with  $T^{(0)}\,=\,0$. In this case,   solutions (\ref{PMfieldequationsolution3}) are unaffected by the functions of the Ricci scalar but   only by  the presence of the Ricci tensor square terms into the interaction Lagrangian. Moreover, from the second line of (\ref{PMfieldequationsolution3}),  we have also that the gravitational waves must be traceless. A prototype of  traceless source is the electromagnetic field where  we have

\begin{eqnarray}\label{PMfieldequationsolution4}
\mathsf{h}^{em}_{\mu\nu}(x)\,=\,\mathcal{X}\,\int d^4x'\mathcal{Z}_2(x,x')\,T^{(0),em}_{\mu\nu}(x')\qquad\text{with}\qquad\mathsf{h}^{em}(x)\,=\,0\,.
\end{eqnarray}
Finally, the solution of the first field Eqs. (\ref{PMfieldequationFOG2}) in the vacuum ($T_{\mu\nu}\,=\,0$) can be derived. By using again the hypothesis of harmonic gauge and the principle of plane wave superposition,  we get

\begin{eqnarray}\label{PMfieldequationsolutionHS}
\mathfrak{h}_{\mu\nu}\,=\,\frac{{m_1}^2-{m_2}^2}
{3\,{m_2}^2}\,\int d^4x'\,\mathcal{Z}_2(x,x')\biggl[
\frac{\partial^2_{\mu'\nu'}}{{m_1}^2}-\frac{\eta_{\mu\nu}}{2}\bigg]\,X^{(1)}_{(hs)}(x')
\nonumber\\\\\nonumber
-2\int d^4x'\mathcal{G}_{GR}(x,x')R^{(1)}_{\mu\nu\,(hs)}(x')+h_{\mu\nu\,(hs)}
\end{eqnarray}
where $X^{(1)}_{(hs)}$, $R^{(1)}_{\mu\nu\,(hs)}$, $h_{\mu\nu\,(hs)}$ are respectively the homogeneous solutions of the Klein-Gordon equation for the Ricci scalar and tensor and  the solution  of wave equation for the metric. The homogeneous solutions are chosen in such a way that they satisfy the boundary conditions and the gauge harmonic condition $\mathfrak{h}^{\mu\rho}_{\,\,\,\,\,\,,\rho}-1/2\,\mathfrak{h}^{\,,\mu}\,=\,0$. Also here, the limit $m\rightarrow 0$  becomes problematic since it is strongly coupled. As above, the remedy is that the sources decouples in this limit \cite{derham}.

An important remark is necessary at this point. In the above calculations, we are using the  harmonic gauge condition that allows to transform admissible field configurations if one considers sufficiently small regions of spacetime. However, such a condition has to work over more than one gravitational wavelengths and the effects of background curvature on wave propagation have to be considered. A detailed discussion  on harmonic gauge condition, shortwave approximation and propagation on curved background can be found in \cite{MTW}. Here we adopted the same approach (see $\S$ 35.13, $\S$ 35.14 therein).

Below we will consider polarizations and helicity states in vacuum starting from this result.

\section{Polarizations and helicity states in vacuum}\label{polar}

We can analyze the  propagation in  vacuum by performing the Fourier analysis of field Eqs. (\ref{PMfieldequationFOG2}). In  the Fourier space, we have

\begin{eqnarray}\label{PMfieldequationFOG2_vacuum}
\begin{array}{ll}
k^2\biggl\{({m_2}^2-k^2)\,\tilde{\mathfrak{h}}_{\mu\nu}-\biggl[\frac{{m_2}^2-{m_1}^2}{3{m_1}^2}\,k_\mu k_\nu+\eta_{\mu\nu}\biggl(\frac{{m_2}^2}{2}-\frac{{m_1}^2+2{m_2}^2}{6{m_1}^2}\,k^2\biggr)\biggr]
\,\tilde{\mathfrak{h}}\biggr\}\,=\,0\\\\
k^2({m_1}^2-k^2)\,\tilde{\mathfrak{h}}\,=\,0
\end{array}
\end{eqnarray}
where $\tilde{\mathfrak{h}}_{\mu\nu}$, $\tilde{\mathfrak{h}}$ are the Fourier representation of $\mathfrak{h}_{\mu\nu}$, $\mathfrak{h}$. The gauge condition now becomes $\tilde{\mathfrak{h}}_{\mu\sigma}\,k^\sigma-1/2\,\tilde{\mathfrak{h}}\,k_\mu\,=\,0$. The solutions are shown in  Table \ref{table_1}.
\begin{table}[ht]
\centering
\begin{tabular}{c|c|c}
\hline\hline\hline
 Case & Choices & Gauge condition \\
 \hline
 A & $\begin{array}{ll}k^2\,=\,0\\\text{any}\,\,\tilde{\mathfrak{h}}_{\mu\nu}\,\text{with}\,\,\tilde{\mathfrak{h}}\,=\,0\end{array}$ & $\begin{array}{ll}
 \tilde{\mathfrak{h}}_{\mu\sigma}\,k^\sigma\,=\,0\end{array}$\\
 \hline
 B & $\begin{array}{ll}k^2\,=\,{m_1}^2\\\tilde{\mathfrak{h}}_{\mu\nu}\,=\,\biggl[\frac{k_\mu k_\nu}{{m_1}^2}+\frac{\eta_{\mu\nu}}{2}\biggr]\frac{\tilde{\mathfrak{h}}}{3}\\\text{with}\,\,\tilde{\mathfrak{h}}\neq 0\,\,\\\end{array}$ & $\begin{array}{ll}
 \text{verified}\end{array}$\\
 \hline
 C & $\begin{array}{ll}k^2\,=\,{m_2}^2\\\text{any}\,\,\tilde{\mathfrak{h}}_{\mu\nu}\,\text{with}\,\,\tilde{\mathfrak{h}}\,=\,0\end{array}$ & $\begin{array}{ll}
 \tilde{\mathfrak{h}}_{\mu\sigma}\,k^\sigma\,=\,0\end{array}$\\
 \hline
 D & $\begin{array}{ll}k^2\,=\,0\\\text{any}\,\,\tilde{\mathfrak{h}}_{\mu\nu}\,\,\text{with}\,\,\tilde{\mathfrak{h}}\neq 0\end{array}$ & $\begin{array}{ll}
 \tilde{\mathfrak{h}}_{\mu\sigma}\,k^\sigma-1/2\,\tilde{\mathfrak{h}}\,k_\mu\,=\,0\end{array}$ \\
 \hline\hline\hline
 \end{tabular}
\caption{\label{table_1}Classification of solutions in the vacuum of field Eqs. (\ref{PMfieldequationFOG2_vacuum}). Specifically, the case A is one of General Relativity}; the case B is massive and pure trace; the case C is massive, trace-free and transverse and the case D is massless where the trace-reverse of $\mathfrak{h}$ is transverse.
\end{table}

The case A corresponds to the standard massless gravitational waves of General Relativity. The solution B is the same of the pure $f(X)$-gravity, \emph{i.e.} the choice $k^2\,=\,{m_1}^2$ reduces automatically the field equations of $f(X,Y,Z)$-gravity to those of $f(X)$-gravity. Cases B and C are very  different. In fact  the solution C is traceless while the solution B satisfies directly the gauge condition.

The solution of case B for $\mathfrak{h}_{\mu\nu}$ is given by the following expression

\begin{eqnarray}\label{GW_B}
\mathfrak{h}^B_{\mu\nu}\,=\,&&\frac{1}
{3}\,\int\frac{d^4k}{(2\pi)^4}\,\biggl[\frac{k_\mu k_\nu}{{m_1}^2}+\frac{\eta_{\mu\nu}}{2}\biggr]\mathfrak{h}^B(k)\,e^{\mathfrak{j}\,kx}\,=\,\frac{1}{3}\biggl[\frac{\eta_{\mu\nu}}
{2}-\frac{\partial^2_{\mu\nu}}
{{m_1}^2}\biggr]\,\int\frac{d^4k}{(2\pi)^4}\,\mathfrak{h}^B(k)\,e^{\mathfrak{j}\,kx}
\nonumber\\\\\nonumber
=&&\frac{1}{3}\biggl[\frac{\eta_{\mu\nu}}
{2}-\frac{\partial^2_{\mu\nu}}
{{m_1}^2}\biggr]\,\mathfrak{h}^B(x)\,,
\end{eqnarray}
where the trace of metric is a generic Klein-Gordon function with $k\,=\,(\omega_1,\,\mathbf{k})$, where $\omega_1\,=\,\sqrt{|\mathbf{k}|^2+{m_1}^2}$. We can write the general solution in terms of its Fourier modes which are plane waves, that is 

\begin{eqnarray}
\mathfrak{h}^B(t,\,\mathbf{x})\,=\,\int\frac{d^3\mathbf{k}}{(2\pi)^3}
\,C(\mathbf{k})\,e^{\mathfrak{j}\,(\omega_1t-\mathbf{k}\cdot\mathbf{x})}
\end{eqnarray}
where $C(\mathbf{k})$ is the Fourier representation of the trace $\mathfrak{h}^B$. If we consider a propagating trace in the $z$-direction\footnote{We set $x^3\,=\,z$.} $\mathfrak{h}^B(t,\,z)\,=\,\mathfrak{h}_0\,e^{\mathfrak{j}\,(\omega_1t-k_z\,z)}$, where $k\,=\,(\omega_1,\,0,\,0,\,k_z)$ with ${\omega_1}^2-{k_z}^2\,=\,{m_1}^2$, the solution (\ref{GW_B}) is given by

\begin{eqnarray}\label{GW_matrix_0}
\mathfrak{h}_{\mu\nu}^B(t,\,z)\,=\,\epsilon^B_{\mu\nu}\,e^{\mathfrak{j}\,(\omega_1 t-k_z\,z)}\,=\,\frac{\mathfrak{h}_0}{3}\begin{pmatrix}
\frac{1}{2}+\frac{{\omega_1}^2}{{m_1}^2}&0&0&-\frac{\omega_1\,k_z}{{m_1}^2}\\
0&-\frac{1}{2}&0&0\\
0&0&-\frac{1}{2}&0\\
-\frac{\omega_1\,k_z}{{m_1}^2}&0&0&-\frac{1}{2}+\frac{{k_z}^2}{{m_1}^2}\\
\end{pmatrix}\,e^{\mathfrak{j}\,(\omega_1 t-k_z\,z)}
\end{eqnarray}
where $\epsilon^B_{\mu\nu}$ is the polarization tensor. By a change of coordinates $x^\mu\,\rightarrow\,x'^\mu\,=\,x^\mu+\zeta^\mu(x)$ with $\mathcal{O}(\zeta^2$)$\ll 1$,  we can transform the metric $\mathfrak{h}_{\mu\nu}$ into a new metric $\mathfrak{h}'_{\mu\nu}\,=\,\mathfrak{h}_{\mu\nu}-\zeta_{\mu,\nu}-\zeta_{\nu,\mu}$ . Let us suppose that we choose $\zeta^\mu(x)\,=\,\mathfrak{j}\,\theta^\mu\,e^{\mathfrak{j}\,kx}$,  the metric $\mathfrak{h}_{\mu\nu}(x)$ becomes $\mathfrak{h}'_{\mu\nu}(x)\,=\,\epsilon'_{\mu\nu}\,e^{\mathfrak{j}\,k\,x}$ where $\epsilon'_{\mu\nu}\,=\,\epsilon_{\mu\nu}+k_\mu\theta_\nu+k_\nu\theta_\mu$. By performing a change of coordinates and choosing $\theta_\mu\,=\,\theta^B_\mu\,=\,(-\frac{1}{4\omega_1}-\frac{\omega_1}{2{m_1}^2},\,0,\,0,\,\frac{k_z}{2{m_1}^2}-\frac{k_z}{4{\omega_1}^2})$ the polarization tensor $\epsilon^B_{\mu\nu}$ in Eq. (\ref{GW_matrix_0}) becomes

\begin{eqnarray}\label{GW_matrix_0_base}
\epsilon'^B_{\mu\nu}\,=\,\frac{\mathfrak{h}_0}{3}\begin{pmatrix}
0&0&0&0\\
0&-\frac{1}{2}&0&0\\
0&0&-\frac{1}{2}&0\\
0&0&0&-\frac{1}{2}+\frac{{k_z}^2}{{2\omega_1}^2}\\
\end{pmatrix}\,=\,-\frac{\mathfrak{h}_0}{6}\begin{pmatrix}
0&0&0&0\\
0&1&0&0\\
0&0&1&0\\
0&0&0&1\\
\end{pmatrix}+\frac{{k_z}^2\mathfrak{h}_0}{{6\,\omega_1}^2}\begin{pmatrix}
0&0&0&0\\
0&0&0&0\\
0&0&0&0\\
0&0&0&1\\
\end{pmatrix}\,.
\end{eqnarray}

In the case C,  we have 

\begin{eqnarray}
\mathfrak{h}^C_{\mu\nu}(t,\,\mathbf{x})\,=\,\int\frac{d^3\mathbf{k}}{(2\pi)^3}
\,C_{\mu\nu}(\mathbf{k})\,e^{\mathfrak{j}\,(\omega_2t-\mathbf{k}\cdot\mathbf{x})}
\end{eqnarray}
where $C_{\mu\nu}(\mathbf{k})$ is the Fourier representation of the gravitational wave $\mathfrak{h}^C_{\mu\nu}$ and $\omega_2\,=\,\sqrt{{|\mathbf{k}|}^2+{m_2}^2}$. Also in this case, by considering a propagating wave in the $z$-direction $\mathfrak{h}_{\mu\nu}^C(t,\,z)\,=\,\epsilon^C_{\mu\nu}\,e^{\mathfrak{j}\,(\omega_2t-k_z\,z)}$, where the polarization tensor $\epsilon^C_{\mu\nu}$ satisfies the traceless condition $\eta^{\rho\sigma}\epsilon^C_{\rho\sigma}\,=\,0$ and after also the harmonic gauge $\epsilon^C_{\mu\sigma}\,k^\sigma\,=\,0$ (in the Fourier space), we find the solution 

\begin{eqnarray}\label{GW_matrix_2}
\mathfrak{h}_{\mu\nu}^C(t,\,z)\,=\,\begin{pmatrix}
\epsilon_{00}&\epsilon_{01}&\epsilon_{02}&-\frac{\omega_2}{k_z}\,\epsilon_{00}\\
\epsilon_{01}&\epsilon_{11}&\epsilon_{12}&-\frac{\omega_2}{k_z}\,\epsilon_{01}\\
\epsilon_{02}&\epsilon_{12}&-\frac{{m_2}^2}{{k_z}^2}\epsilon_{00}-\epsilon_{11}&-\frac{\omega_2}{k_z}\,\epsilon_{02}\\
-\frac{\omega_2}{k_z}\,\epsilon_{00}&-\frac{\omega_2}{k_z}\,\epsilon_{01}&-\frac{\omega_2}{k_z}\,\epsilon_{02}&\frac{{\omega_2}^2}{{k_z}^2}\,\epsilon_{00}\\
\end{pmatrix}
\,e^{\mathfrak{j}\,(\omega_2 t-k_z z)}
\end{eqnarray}
where $\epsilon_{00}$, $\epsilon_{01}$, $\epsilon_{02}$, $\epsilon_{11}$, $\epsilon_{12}$  are unspecified values. By performing also in this case a change of coordinates \cite{greci1,greci2} and choosing $\theta_\mu\,=\,\theta^C_\mu\,=\,(-\frac{\epsilon_{00}}{2\omega_2},\,-\frac{\epsilon_{01}}{\omega_2},\,-\frac{\epsilon_{02}}{\omega_2},\,\frac{\epsilon_{00}}{k_z}-\frac{k_z\epsilon_{00}}{2{\omega_2}^2})$ the polarization tensor $\epsilon^C_{\mu\nu}$ in Eq.(\ref{GW_matrix_2}) becomes

\begin{eqnarray}\label{GW_matrix_2_base}
\epsilon'^C_{\mu\nu}\,=\,&&\begin{pmatrix}
0&0&0&0\\
0&\epsilon_{11}&\epsilon_{12}&-\frac{{m_2}^2}{k_z\omega_2}\,\epsilon_{01}\\
0&\epsilon_{12}&-\frac{{m_2}^2}{{k_z}^2}\epsilon_{00}-\epsilon_{11}&-\frac{{m_2}^2}{k_z\omega_2}\,\epsilon_{02}\\
0&-\frac{{m_2}^2}{k_z\omega_2}\,\epsilon_{01}&-\frac{{m_2}^2}{k_z\omega_2}\,\epsilon_{02}&\frac{{m_2}^4}{{k_z}^2{\omega_2}^2}\,\epsilon_{00}\\
\end{pmatrix}\,=\,\epsilon_{11}\begin{pmatrix}
0&0&0&0\\
0&1&0&0\\
0&0&-1&0\\
0&0&0&0\\
\end{pmatrix}+\epsilon_{12}\begin{pmatrix}
0&0&0&0\\
0&0&1&0\\
0&1&0&0\\
0&0&0&0\\
\end{pmatrix}\nonumber\\\\\nonumber\\&&-\frac{{m_2}^2\epsilon_{00}}{{k_z}^2}\begin{pmatrix}
0&0&0&0\\
0&0&0&0\\
0&0&1&0\\
0&0&0&0\\
\end{pmatrix}+\frac{{m_2}^4\epsilon_{00}}{{k_z}^2{\omega_2}^2}\begin{pmatrix}
0&0&0&0\\
0&0&0&0\\
0&0&0&0\\
0&0&0&1\\
\end{pmatrix}-\frac{{m_2}^2\,\epsilon_{01}}{k_z\omega_2}\begin{pmatrix}
0&0&0&0\\
0&0&0&1\\
0&0&0&0\\
0&1&0&0\\
\end{pmatrix}\,\,-\frac{{m_2}^2\epsilon_{02}}{k_z\omega_2}\begin{pmatrix}
0&0&0&0\\
0&0&0&0\\
0&0&0&1\\
0&0&1&0\\
\end{pmatrix}\nonumber\,.
\end{eqnarray}

Finally for the case D we have

\begin{eqnarray}\label{GW_matrix_3}
\mathfrak{h}_{\mu\nu}^D(t,\,z)\,=\,\begin{pmatrix}
\frac{\epsilon}{2}-\epsilon_{03}&\epsilon_{01}&\epsilon_{02}&\epsilon_{03}\\
\epsilon_{01}&\epsilon_{11}&\epsilon_{12}&-\epsilon_{01}\\
\epsilon_{02}&\epsilon_{12}&-\epsilon_{11} &-\epsilon_{02}\\
\epsilon_{03}&-\epsilon_{01}&-\epsilon_{02}&-\frac{\epsilon}{2}-\epsilon_{03}\\
\end{pmatrix}
\,e^{\mathfrak{j}\,\omega\,(t-z)}
\end{eqnarray}
where $\epsilon_{01}$, $\epsilon_{02}$, $\epsilon_{03}$, $\epsilon_{11}$, $\epsilon_{12}$ are unspecified values and $\epsilon$ is the trace of polarization tensor $\epsilon^D_{\mu\nu}$. Here if we choose $\theta_\mu\,=\,\theta^D_\mu\,=\,(\frac{2\epsilon_{03}-\epsilon}{4\omega},\,-\frac{\epsilon_{01}}{\omega},\,-\frac{\epsilon_{02}}{\omega},\,-\frac{2\epsilon_{03}+\epsilon}{4\omega})$ the polarization tensor $\epsilon^D_{\mu\nu}$ in Eq.(\ref{GW_matrix_3}) becomes

\begin{eqnarray}\label{GW_matrix_3_base}
\epsilon'^D_{\mu\nu}\,=\,&&\begin{pmatrix}
0&0&0&0\\
0&\epsilon_{11}&\epsilon_{12}&\epsilon_{01}+\epsilon_{13}\\
0&\epsilon_{12}&-\epsilon_{11} &\epsilon_{02}+\epsilon_{23}\\
0&\epsilon_{01}+\epsilon_{13}&\epsilon_{02}+\epsilon_{23}&0\\
\end{pmatrix}\,=\,\epsilon_{11}\begin{pmatrix}
0&0&0&0\\
0&1&0&0\\
0&0&-1&0\\
0&0&0&0\\
\end{pmatrix}+\epsilon_{12}\begin{pmatrix}
0&0&0&0\\
0&0&1&0\\
0&1&0&0\\
0&0&0&0\\
\end{pmatrix}\nonumber\\\\\nonumber\\&&+(\epsilon_{01}+\epsilon_{13})\begin{pmatrix}
0&0&0&0\\
0&0&0&1\\
0&0&0&0\\
0&1&0&0\\
\end{pmatrix}+(\epsilon_{02}+\epsilon_{23})\begin{pmatrix}
0&0&0&0\\
0&0&0&0\\
0&0&0&1\\
0&0&1&0\\
\end{pmatrix}\nonumber\,.
\end{eqnarray}
{However the last two polarizations in (\ref{GW_matrix_3_base}) are not physical and pratically we obtained the same outcome of General Relativity (case A).

If we introduce an independent basis (see Table \ref{polarization}) for the polarizations (\ref{GW_matrix_0_base}) and (\ref{GW_matrix_2_base}),  the general solution of field Eqs. (\ref{PMfieldequationFOG2}) in  vacuum is 

\begin{eqnarray}\label{GW_polarization}
\mathfrak{h}_{\mu\nu}(t,\,z)\,=\,&&\biggl[\mathcal{H}_1\,\epsilon^{(+)}_{\mu\nu}+\mathcal{H}_2\,\epsilon^{(\times)}_{\mu\nu}\biggr]\,e^{\mathfrak{j}\,k_z(t-z)}+\mathcal{H}_3\biggl[\epsilon^{(1)}_{\mu\nu}-\frac{{k_z}^2}{\sqrt{3}\,{\omega_1}^2}\epsilon^{(S)}_{\mu\nu}\biggr]\,e^{\mathfrak{j}\,(\omega_1t-k_z\,z)}
\nonumber\\\nonumber\\
&&+\biggl[\mathcal{H}_4\,\epsilon^{(+)}_{\mu\nu}+\mathcal{H}_5\,\epsilon^{(\times)}_{\mu\nu}+\mathcal{H}_6\biggl(\sqrt{3}\,\epsilon^{(1)}_{\mu\nu}-\sqrt{2}\,\epsilon^{(+)}_{\mu\nu}-\frac{{\omega_2}^2+2\,{m_2}^2}{{\omega_2}^2}\,\epsilon^{(S)}_{\mu\nu}\biggr)
\\\nonumber\\\nonumber
&&\,\,\,+\mathcal{H}_7\,\epsilon^{(2)}_{\mu\nu}+\mathcal{H}_8\,\epsilon^{(3)}_{\mu\nu}\biggr]\,e^{\mathfrak{j}\,(\omega_2t-k_z\,z)}
\end{eqnarray}
where $\mathcal{H}_1$, $\mathcal{H}_2$ are arbitrary constants related  to the propagation modes of gravitational waves in General Relativity  and the other constants are defined as  $\mathcal{H}_3\,=\,-\frac{\sqrt{3}\,\mathfrak{h}_0}{6}$, $\mathcal{H}_4\,=\,\epsilon_{11}$, $\mathcal{H}_5\,=\,\epsilon_{12}$, $\mathcal{H}_6\,=\,-\frac{{m_2}^2\epsilon_{00}}{2{k_z}^2}$, $\mathcal{H}_7\,=\,-\frac{\sqrt{2}\,{m_2}^2\epsilon_{01}}{k_z\omega_2}$, $\mathcal{H}_8\,=\,-\frac{\sqrt{2}\,{m_2}^2\epsilon_{02}}{k_z\omega_2}$.

\begin{table}[ht]
\centering
\begin{tabular}{ccc}
\hline\hline\hline
&&\\
$\epsilon^{(+)}_{\mu\nu}\,=\,\frac{1}{\sqrt{2}}\begin{pmatrix}
0&0&0&0\\
0&1&0&0\\
0&0&-1&0\\
0&0&0&0\\
\end{pmatrix}$ & $\epsilon^{(\times)}_{\mu\nu}\,=\,\frac{1}{\sqrt{2}}\begin{pmatrix}
0&0&0&0\\
0&0&1&0\\
0&1&0&0\\
0&0&0&0\\
\end{pmatrix}$ & $\epsilon^{(S)}_{\mu\nu}\,=\,\begin{pmatrix}
0&0&0&0\\
0&0&0&0\\
0&0&0&0\\
0&0&0&1\\
\end{pmatrix}$ \\
&&\\
$ \epsilon^{(1)}_{\mu\nu}\,=\,\frac{1}{\sqrt{3}}\begin{pmatrix}
0&0&0&0\\
0&1&0&0\\
0&0&1&0\\
0&0&0&1\\
\end{pmatrix}$ & $\epsilon^{(2)}_{\mu\nu}\,=\,\frac{1}{\sqrt{2}}\begin{pmatrix}
0&0&0&0\\
0&0&0&1\\
0&0&0&0\\
0&1&0&0\\
\end{pmatrix}$ & $\epsilon^{(3)}_{\mu\nu}\,=\,\frac{1}{\sqrt{2}}\begin{pmatrix}
0&0&0&0\\
0&0&0&0\\
0&0&0&1\\
0&0&1&0\\
\end{pmatrix}$\\
 &&\\
 \hline\hline\hline
 \end{tabular}
\caption{\label{polarization}The basis of the polarizations. Each polarization satisfies the condition $\epsilon_{\rho\sigma}\epsilon^{\rho\sigma}\,=\,1$.}
\end{table}

The different components of  polarization tensor can be distinguished if we ask how $\epsilon_{\mu\nu}$ changes when  the coordinate system undergoes  a rotation of a given  angle $\varphi$ about the $z$-axis \cite{weinberg}. This is a  Lorentz transformation of the form

\begin{eqnarray}\label{rotation}
\mathcal{R}_\mu^{\,\,\,\,\nu}\,=\,\begin{pmatrix}
1&0&0&0\\
0&\cos\varphi&\sin\varphi&0\\
0&-\sin\varphi&\cos\varphi&0\\
0&0&0&1\\
\end{pmatrix}
\end{eqnarray}
and since it leaves $k^\mu$ invariant ($\mathcal{R}_\mu^{\,\,\,\,\sigma}\,k_\sigma\,=\,k_\mu$), the only effect is to transform $\epsilon_{\mu\nu}$ into $\tilde{\epsilon}_{\mu\nu}\,=\,\mathcal{R}_\mu^{\,\,\,\,\rho}\,\mathcal{R}_\nu^{\,\,\,\,\sigma}\,\epsilon_{\rho\sigma}$. In the case B,  the polarization tensor $\epsilon'^B_{\mu\nu}$ is unchanged, then we can state that the helicity is null\footnote{Any plane wave $\psi$ transforming  under a rotation of an angle $\varphi$ about the direction of propagation into $\tilde{\psi}\,=\,e^{\mathfrak{j}\,\xi\,\varphi}\psi$  has helicity $\xi$.}. In the case C,  we have

\begin{eqnarray}\label{helicity}
\tilde{\epsilon}'^C_{\pm}\,&=&\,e^{\pm \mathfrak{j}\,2\,\varphi}\,\epsilon'^C_{\pm}+\frac{{m_2}^2\epsilon_{00}}{{k_z}^2}\sin\varphi\,e^{\pm\,\mathfrak{j}\,(\varphi+\pi/2)}\nonumber\\\\\nonumber
\tilde{l}'^C_{\pm}\,&=&\,e^{\pm \mathfrak{j}\,\varphi}\,l'^C_{\pm}
\end{eqnarray}
where $\epsilon_{\pm}\,\doteq\,\epsilon_{11}\mp \mathfrak{j}\,\epsilon_{12}$ and $l_{\pm}\,\doteq\,\epsilon_{13}\mp \mathfrak{j}\,\epsilon_{23}$. The helicity is $\pm\,1$ for the states $l_\pm$, while for $\epsilon_{\pm}$ {we cannot identify the condition $\tilde{\epsilon}'^C_{\pm}\,=\,e^{\pm \mathfrak{j}\,\xi\,\varphi}\,\epsilon'^C_{\pm}$. As such condition holds only if $\epsilon_{00}\,\neq\,0$ (and $\mathcal{H}_6\,\neq\,0$ in Eq.(\ref{GW_polarization})). In other words, we  have the possible helicity values  $0,\,\pm 1,\,\pm 2$. Clearly, all the polarization tensors discussed here can be referred to the Wigner little group $E(2)$ as discussed in \cite{eardley}.

\section{Conclusions}\label{conclusions}

In this paper, we presented a complete study of post-Minkowskian limit and gravitational wave solutions of fourth-order gravity theories of gravity in four dimensional space-time. We considered a generic action constructed with curvature invariants derived from the Riemann tensor, that are the Ricci curvature scalar, the squared Ricci tensor and the squared Riemann tensor. Thanks to the Gauss-Bonnet topological invariant, it is possible to consider only two curvature invariants since the third is always related to the others by the Gauss-Bonnet constraint. In this sense, all the budget of degrees of freedom can be related to $R$ and $R_{\alpha\beta}$. 

With respect to standard General Relativity, new features come out in the post-Minkowskian limit. First of all two further massive scalar modes emerge in relation to the non linearity in the Ricci scalar and tensor terms. This means that massive gravitons are a  characteristic of this theories and their effective masses are strictly related to the form of the action $f$. As supposed by several authors, these particles could play a fundamental role for  the dark matter issue that, in this case, could directly come from the gravitational part of cosmic dynamics \cite{graviton1,graviton2,attila}. 

Furthermore, the total number of polarizations is six and helicity can come into three distinct states. It is worth noticing that we have not to choose arbitrarily $\times$ and $+$ polarizations as in General Relativity but all possible polarizations naturally come out. This fact could be of great  interest for the gravitational wave detection since running and forthcoming experiments  could take advantage from this theoretical result and investigate new scenarios. In a forthcoming paper, a detailed study of sources compatible with these results will be pursued.

\section*{Acknowledgements}
SC acknowledges INFN Sez. di Napoli (Iniziative Specifiche QGSKY, and TEONGRAV) for financial support.

\end{document}